# Overlap Technique for End-Cap Seals on Cylindrical Magnetic Shields

S. Malkowski[1], R.Y. Adhikari[1], J. Boissevain[2], C. Daurer[3], B.W. Filippone[2], B. Hona[1], B. Plaster[1], D. Woods[1], and H. Yan[1]

[1]Department of Physics and Astronomy, University of Kentucky, Lexington, KY 40506 USA
[2]Kellogg Radiation Laboratory, California Institute of Technology, Pasadena, CA 91125 USA
[3]Bartoszek Engineering, Aurora, IL 60506 USA

We present results from studies of the effectiveness of an overlap technique for forming a magnetic seal across a gap at the boundary between a cylindrical magnetic shield and an end-cap. In this technique a thin foil of magnetic material overlaps the two surfaces, thereby spanning the gap across the cylinder and the end-cap, with the magnetic seal then formed by clamping the thin magnetic foil to the surfaces of the cylindrical shield and the end-cap on both sides of the gap. In studies with a prototype 31-cm diameter, 91-cm long, 0.16-cm thick cylindrical magnetic shield and flared end-cap, the magnetic shielding performance of our overlap technique is comparable to that obtained with the conventional method in which the end-cap is placed in direct lapped contact with the cylindrical shield via bolts or screws.

## I. INTRODUCTION

THE conventional technique for forming a magnetic seal between an open-ended cylindrical magnetic shield and an end-cap (consisting of either a flat or cone-shaped/flared surface with a short cylindrical section having an inner diameter slightly larger than the open-ended cylindrical shield's outer diameter) is to overlap the end-cap with the cylindrical shield, and then directly couple the two surfaces via through bolts or screws (thus, requiring small penetrations through the cylinder and end-cap surfaces) located at repeated points along the circumference. In order for this technique to be effective, the end-cap typically overlaps the cylindrical shield; thus, this technique can be quite sensitive to manufacturing tolerances on the cylinder and end-cap's roundness and diameter on large (meter-scale) shields (e.g., if large radial gaps persist after overlapping). This can potentially be problematic for large-scale cylindrical magnetic shields (i.e., diameters on the order of a meter or larger), which are typically assembled from smaller individually-fabricated subsections manufactured according to standard sheet-metal techniques (generally, the size of which are constrained by the manufacturer's annealing oven). Further, there can be a significant labor burden associated with the assembly or disassembly of the end-cap on such a large-scale cylindrical shield as the through bolts or screws are typically spaced approximately every 10–15 cm along the circumference. The use of bolts or screws also presents an increased potential for damage from, for example, over-tightening of the bolts or screws (thus stressing the magnetic shield material) or from accidents (such as dropped tools).

In an alternative technique presented in the literature [1], a flat end-cap is coupled to a wall flange on the cylindrical shield, and the seal is then formed via through bolts and clamp rings on the interior flange surface and the exterior end-cap surface. The merit of this technique is that it is less sensitive to the relative diameters and roundness of the cylindrical shield and the end-cap.

In the remainder of this article we present another alternative technique, which we term an "overlap technique", for forming a magnetic seal between a cylindrical shield and an end-cap. We developed this overlap technique with the idea of applying it to large-scale shields (diameters on the scale of several meters, lengths on the scale of tens of meters), where it may not be possible to control the tolerances such that there are no gaps between the cylindrical shield and the end-cap along the large-scale circumferences. A thin sheet of magnetic foil overlaps the two surfaces, thereby spanning any such gaps at the boundary between the cylindrical shield and the end-cap, and a continuous circumferential seal is then formed by clamping this thin foil to the cylindrical shield and the end-cap on both sides of the gap. The labor burden required during the assembly or disassembly of the end-cap on the cylindrical shield is also reduced, because the seal is formed with continuous clamp rings, as opposed to many through bolts or screws.

## II. OVERLAP TECHNIQUE

A schematic diagram of our overlap technique is shown in Fig. 1, and the primary features are as follows. Aluminum backer rings are attached to the interior surfaces of the cylindrical shield and the end-cap. A thin magnetic foil spans the gap at the boundary between the cylindrical shield and the end-cap along the length of the circumference. Stainless steel clamp rings positioned on the exterior surfaces of the cylindrical shield and the end-cap are tightened against the backer rings. This secures the thin magnetic foil to the cylindrical shield and the end-cap surfaces, thereby forming the magnetic seal across the boundary.

An additional feature of our overlap technique is that the backer rings on the cylindrical shield and the end-cap can be designed so that they protrude slightly beyond the ends of the magnetic shield material. These protruding backer rings then effectively function as "bumpers", providing a protective buffer gap between the magnetic shield material on the cylindrical shield and the end-cap. This can be an especially important consideration on large-scale (i.e., several-meter diameter) magnetic shields, where there is the potential for damage during the positioning of the end-cap with machinery.

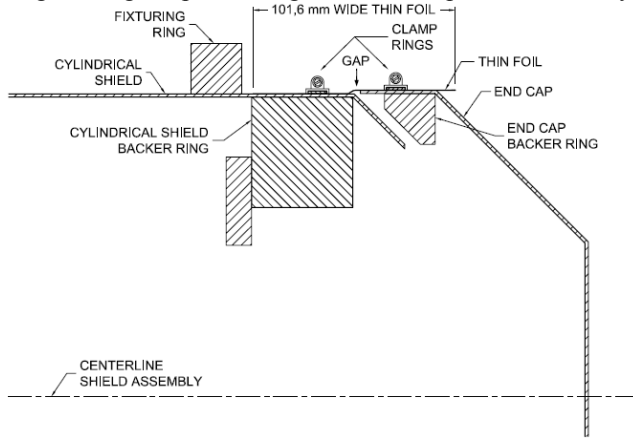

Figure 1: Schematic diagram of the overlap technique.

## III. PROTOTYPE

We tested our overlap technique method with a prototype magnetic shield system consisting of a 31-cm diameter, 91-cm long, 0.16-cm thick "$\mu$-metal" cylindrical shield and a removable flared end-cap [2]. Note that this shield was originally designed and manufactured such that this removable end-cap was to be coupled to the cylinder via the "conventional technique" described in the Introduction to this article. In particular, the end-cap inner diameter is slightly larger than the cylinder outer diameter, thus permitting the end-cap to directly overlap the cylinder. Small (~0.5 cm) holes through the end-cap and the cylinder, located at 12-cm intervals around the circumference, permitted direct coupling of the two overlapped surfaces via screws, with nuts welded onto the interior surface of the cylinder. The opposite end of the cylinder included a fixed conical-shaped end-cap with a 5.08-cm diameter central hole (which permitted measurements of the residual fields inside the shield when the removable end-cap was installed).

The shield was then retrofitted in order to test our overlap technique. Photographs of the cylinder and the end-cap with their respective aluminum backer rings installed and also a three-dimensional cutaway CAD rendering of this prototype system are shown in Fig. 2. In order to install the backer ring inside the cylinder, the ring was split into four azimuthal sections, and then assembled in-situ into a contiguous ring via couplings to an aluminum connector plate disc. The backer ring installed inside of the flared end-cap was designed such that it also functioned as a stand-off, so that there was a small gap (~2 mm) between the cylinder and end-cap surfaces in the axial direction.

For the overlapping thin magnetic foil, we employed 10.16-cm wide CO-NETIC AA foils [3] in two different thicknesses: 0.00508-cm and 0.01524-cm. We chose this particular foil from this vendor based on its permeability characteristics in external fields of magnitude similar to the Earth's field and because the available width was well matched to spanning a gap with sufficient overlap for attachment of the clamp rings on both sides of the gap; foils of similar permeability and width from other vendors would likely yield similar results. Figure 3 shows a photograph of the prototype before and after the foil was secured in place with the clamp rings.

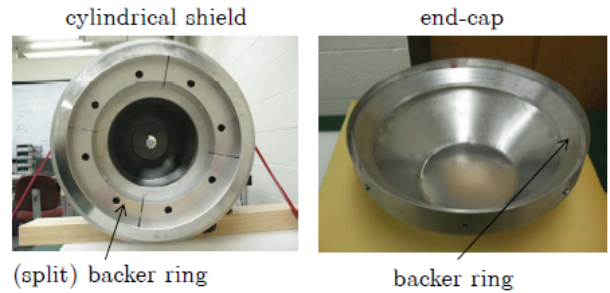

Figure 2: Photographs of the prototype cylindrical shield (top left) and the removable end-cap (top right) with their backer rings installed. A three-dimensional cutaway CAD rendering of the prototype system is shown in the bottom figure.

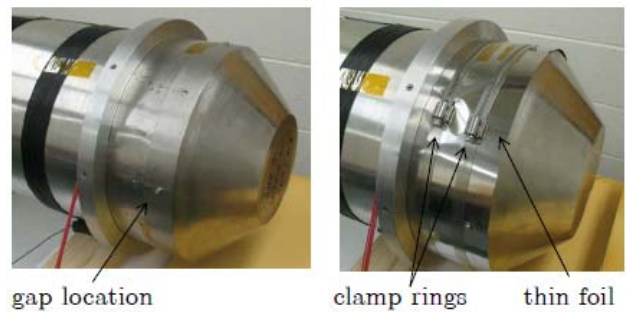

Figure 3: Photographs of the prototype before (left) and after (right) the foil was secured in place with stainless steel clamp rings.

## IV. MEASUREMENTS AND RESULTS

In order to test our prototype system, we measured the residual magnetic fields inside of the magnetic shield with an automated magnetic mapping system which consisted of a computer-controlled, three-axis stepper motor assembly [4]. This system controlled the movement of a low-noise triple-axis fluxgate magnetometer (with a resolution better than ±10 μGauss), which was mounted on the end of a 210-cm long non-magnetic G10 arm. The probe was inserted through the 5.08-cm central hole in the fixed end-cap. All of the measurements were conducted after a 60 Hz AC degaussing cycle. The ambient background field consisted of a 0.45 Gauss transverse field and a 0.16 Gauss axial field (axial and transverse directions defined relative to the shield axis).

Results from our measurements of the residual axial and transverse fields are shown in Fig. 4 for the different end-cap scenarios. There we compare the residual fields for the "conventional technique" (i.e., the end-cap attached directly to the cylindrical shield via through screws), the overlap technique with the 0.00508-cm thick CO-NETIC AA foil and the 0.01524-cm thick CO-NETIC AA foil, the case where the end-cap was in place but with no overlap foil, and the case of no end-cap present on the shield. As can be seen there, the performance of the overlap technique is comparable to, if not better than, the conventional technique in the central region and in the region approaching the end-cap. Also, we do note that the performance for the case where the end-cap was in place but with no overlap foil was inferior to the case where the foil was in place and secured via the overlap technique.

The performance of the thicker 0.01524-cm CO-NETIC AA foil was somewhat better than the thinner foil. This slightly superior performance of the thicker foil is not surprising based on the usual magnetic shielding expectation that the magnetic shielding factor scales linearly with thickness [5]. In addition, we also note a practical consideration that the thicker 0.01524-cm foil was "stiffer", and thus less susceptible to "crinkling" during the tightening of the ring clamps.

In addition, we conducted additional tests in which we forced the gap between the cylindrical shield and the end-cap to be ~1.27 cm. The performance of our overlap technique (with both the 0.00508-cm and 0.01524-cm thick CO-NETIC AA foils) was again comparable to the conventional technique, demonstrating that ~1.27 cm gaps between the cylindrical shield and an end-cap can be included in the engineering design of large-scale magnetic shields. We show results from these tests for the residual axial fields in Fig. 5 (the results for the residual transverse fields were similar to those shown in Fig. 4). In comparing Fig. 5 with Fig. 4, it can be seen that for the case of the end-cap in place but with no overlap foil, the performance is worse (as would be expected) for the larger (1.27 cm) gap. However, the data show that after inclusion of the overlap foil, the axial shielding for the test with the larger gap is more or less identical to that from the test with the smaller gap.

Finally, as a benchmark for the overall performance of our shield, using the standard formula [5] for the transverse shielding factor of a single-shell cylindrical shield with radius $R$ (= 15.5 cm), length $L$ (= 91 cm), thickness $t$ (= 0.16 cm), and relative permeability $\mu$, $S_T = \mu t/2R$, we extract a relative permeability of $\mu \sim 8 \times 10^5$ for our external transverse background fields of ~0.45 G and a residual transverse field of ~0.0001 G, which is consistent (within a factor of two) of the manufacturer's stated maximum permeability of $4 \times 10^5$ [2].

## V. SUMMARY

To conclude, we have demonstrated an overlap technique for the formation of a magnetic seal between a cylindrical magnetic shield and an end-cap. In our technique, a thin sheet of magnetic foil is used to span a gap at the boundary between a cylindrical magnetic shield and an end-cap. The magnetic seal is then formed by clamping this thin foil with clamp rings to the cylinder and end-cap surfaces on both sides of the gap. The performance of this overlap technique was shown to be comparable to, and perhaps better than, the performance obtained under a more conventional technique in which the end-cap was placed in direct contact with the cylinder, and then secured to the cylinder via through screws.

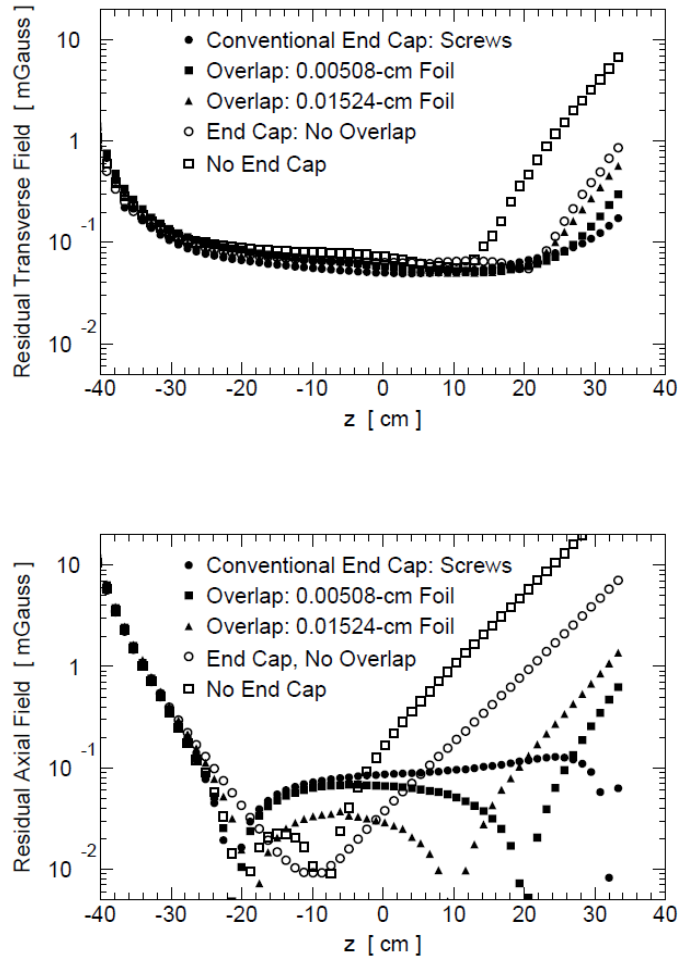

Figure 4: Results from measurements of the residual transverse fields (top panel) and axial fields (bottom panel) for the different end-cap scenarios. The coordinate system is such that the center of the cylindrical shield is at z = 0, and the removable end-cap is at z = +45.5 cm. The magnitude (absolute values) of the transverse and axial fields are plotted; thus, the "kinks" in the axial fields correspond to sign changes (zero crossings) of the axial field component.

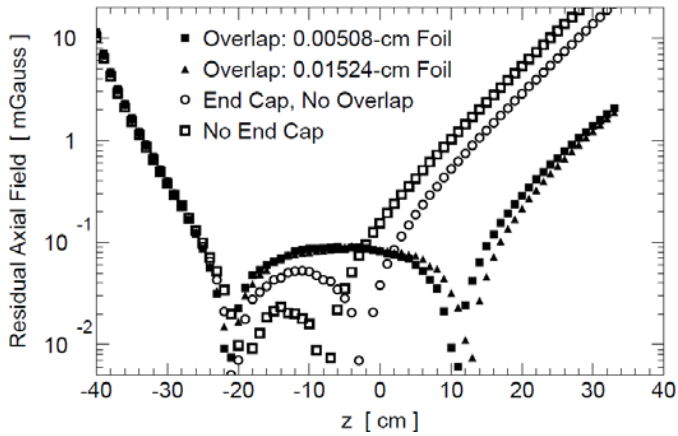

Figure 5: Results from measurements of the residual axial fields for the different end-cap scenarios for the case where the gap between the cylindrical shield and the end-cap was forced to be ~1.27 cm. The coordinate system is the same as in Fig. 4, and the magnitude (absolute value) of the residual axial field is again plotted here.

ACKNOWLEDGMENTS

This work was supported in part by the U.S. Department of Energy Office of Nuclear Physics under Award No. DE-FG02-08ER41557, and by the University of Kentucky. We thank two anonymous referees for their useful comments.